.........................................................................
\magnification=1200
\baselineskip=20pt

\def\delgbl{\delta g^b_L}
\def\rb{R_b}
\def\delrb{\delta R_b}

\def\zl{Z_L}
\def\cw{c_w}
\def\mt{m_t}
\def\delftl{\delta F^t_L}
\def\delftr{\delta F^t_R}
\def\gmmuu{\gamma^{\mu}}
\def\gmmud{\gamma_{\mu}}
\def\ftlsm{F^t_{L, sm}}
\def\delftv{\delta F^t_v}
\def\delfta{\delta F^t_a}

\centerline{\bf Non-commuting ETC corrections to $zt\bar{t}$ vertex}
\vskip .4truein
\centerline{\bf Uma Mahanta}
\centerline{\bf Mehta Research Institute}
\centerline{\bf 10 Kasturba Gandhi Marg}
\centerline{\bf Allahabad-211002}
\centerline{\bf India}
\vskip 1truein
\centerline{\bf Abstract}

In this paper we calculate the corrections to $zt\bar {t}$ couplings
induced by non-commuting ETC interactions. The extra parameters 
$\delgbl$, $s^2$ and ${1\over x}$ of non-commuting ETC models
and the usual SM parameters
will be be assumed to be determined from a global fit to the LEP 1
EW data. We find that  
in the  heavy  (light) case $F^t_L$ is modified by at most 
4\% (2.8\%) relative to its SM value
provided $\delrb < .0088$. This implies that it will be very
difficult to disentangle the ETC corrections from SM corrections
to $zt\bar {t}$ vertex or to probe the 
effects of non-commuting ETC
on $zt\bar {t}$  couplings with the projected  NLC precision
of measuring them.
\vfill
\eject

\centerline{\bf I. Introduction}
      
In ETC models the large mass of the top quark is presumably [1] due to
a low enough
ETC scale ($\approx$ 1 Tev). The sideways ETC interaction that gives
 rise to the large
mass for the top quark also gives rise to a sizeable negative correction
[2] to
$\rb$. This result was considered to be disastrous for ETC models since
the LEP value for $\rb$ at that time was already 
2.2$\sigma$ above the SM prediction.
In order to resolve the $\rb$ anomaly in the context of ETC models 
Chivukula, Simmons and Terning (CST) proposed the non-commuting ETC models
[3] in which the electroweak $SU(2)_h$ gauge group for the 
heavy third generation
 fermions is embedded in the ETC gauge group. The 
 electroweak $SU(2)_l$ gauge interactions of the fermions belonging to the
 first two generations were however 
 assumed to commute with the ETC interactions as usual.
  The same authors also
showed that although these models
contain three extra parameters
($\delgbl , s^2\  and\  {1\over x}$), they provide a better fit [4] to the 
LEP 1 EW data than the SM particularly if $\alpha_s(M^2_z)\approx .112$
as suggested by the low energy deep inelastic scattering
 experiments.
LEP 1 data however does not probe the $zt\bar {t}$ couplings,
the corrections to which
can vary from one ETC model to another even for a fixed $\delrb$.
Precision study of $zt\bar {t}$ coupling can therefore complement our 
understanding of third generation flavor physics derived from
LEP 1 data on $zb\bar {b}$ coupling.
In order to probe non-commuting ETC effects in the process $e^+e^-
\rightarrow t\bar {t}$ far below the $Z_H$ resonance one has to depend
on precision measurements of $zt\bar {t}$ couplings at NLC 500 or higher.
The aim of this article is  to calculate the corrections
to  $zt\bar {t}$ couplings for non-commuting ETC models 
assuming $\rb$ to lie around its present LEP 1 value ($\rb=.2178\pm.0011$)
[5] and to compare
them with the projected NLC sensitivity for measuring them. 

In non-commuting ETC models one has to go through a sequence of symmetry
 breakings- $G_{etc}\times SU(2)_l\times U(1)^{\prime}\rightarrow 
 G_{tc}\times SU(2)_h\times SU(2)_l\times U(1)_y\rightarrow G_{tc}\times
 SU(2)_L\times U(1)_y\rightarrow G_{tc}\times U(1)_Q $-
 in order to give masses to EW gauge bosons and ordinary fermions. 
 To implement the symmetry breaking $SU(2)_h\times SU(2)_l
\times U(1)_y\rightarrow SU(2)_L\times U(1)_y\rightarrow U(1)_Q $ 
 one needs two order parameters
$<\sigma >$ and $<\phi >$. 
$<\sigma >$ breaks $SU(2)_h\times SU(2)_l$ into $SU(2)_L$
and $<\phi >$ breaks $SU(2)_L\times U(1)_y$ into $U(1)_Q$.
Two simplest possibilities of $SU(2)_h\times SU(2)_l$ transformation
 properties of $<\sigma >$ and $<\phi >$ that produce the correct
 mixing and breaking of the gauge groups are:  a) $<\phi >=
 (2, 1)_{1\over 2}$, $<\sigma >=(2, 2)_0$ (heavy case) and
 b)$<\phi >=
 (1, 2)_{1\over 2}$, $<\sigma >=(2, 2)_0$ (light case). In the heavy (light)
 case $<\phi >$ transforms non-tivially under $SU(2)_h (SU(2)_l)$.
 The heavy case corresponds to the situation where the symmetry 
 breaking mechanism that gives mass to the top quark also provides the
 bulk of EW symmetry breaking as evidenced in W and Z boson masses.
 On the other hand in the light case the physics responsible for
 top quark mass does not provide the bulk of W and Z boson masses.
Corrections to $zb\bar {b}$ and $zt\bar {t}$ couplings for non-commuting 
ETC models can arise from two sources: i) sideways ETC induced vertex
 correction and ii) $Z_1-Z_2$ mass mixing.
 
  i)Effect of sideways gauge boson exchange: In non-commuting ETC
  models the sideways ETC gauge boson exchange induces the following
coupling [4] between the Z boson and the LH t-b doublet.

$$\delta L^s_{4f}\approx {g\over \cw}{\xi^2 f^2_Q\over 2f^2}
\bar{\Psi}_L\gmmuu\Psi_L Z_{\mu}.\eqno(1)$$

This implies that $\delta g^{bs}_L=-{\xi^2 f^2_Q\over 2f^2}$ and
$\delta g^{ts}_L=-{\xi^2 f^2_Q\over 2f^2}$ where we have divided 
out by a common factor of $-{g\over \cw}$. In the above 
$f=2 {M_{etc}\over g_{etc}}$ is the scale
at which the ETC gauge group breaks down. In the weak perturbative 
realization of ETC gauge interactions $M_{etc}$ is the mass of
the ETC gauge boson  and $g_{etc}$ is the ETC gauge coupling.

ii)Effect of $Z_1-Z_2$ mass mixing: In non-commuting ETC models
the effect of $Z_1-Z_2$ mass mixing on $\delta g^{b}_L$ and 
$\delta g^{t}_L$ depend on whether the symmetry breaking
pattern correponds to heavy case or light case. We shall
discuss them separately in the following.

a) Heavy case: In the heavy case the $Z_1-Z_2$ mass mixing gives rise to
the following [4] mass eigenstates: $Z_1\approx Z_L-{cs^3\over x\cw}Z_H$
and $Z_2\approx Z_H+{cs^3\over x\cw}Z_L$. Here $Z_L$ and $Z_H$ are mass
eigenstates.
 $Z_L$ wiil be identified with the light neutral Z boson
of the SM. $\phi $ is the angle that characterizes the mixing between
$SU(2)_h$ and $SU(2)_l$.
 So in the heavy case the mixing effect gives rise to the
following coupling between $Z_L$ and the LH (t-b) doublet:
$$\delta L^m= {g\over \cw}{s^4\over x}\bar {\Psi}_L\gmmuu T_{3h}\Psi _L
\zl^{\mu}.\eqno(2)$$
 It then follows that $\delta g^{bm}_L={s^4\over 2x}$ and 
$\delta g^{tm}_L=-{s^4\over 2x}$. The overall correction to the $zb\bar{b}$
and $zt\bar{t}$ couplings in the heavy case are given by $\delta g^b_L=
-{\xi^2 f^2_Q\over 2f^2}+{s^4\over 2x}$ and 
$\delta g^t_L=-{\xi^2 f^2_Q\over 2f^2}-{s^4\over 2x}$. We find that in the
heavy case the effects due to mixing and sideways gauge boson induced 
vertex correction interfere constructively (destructively) in LH
$zt\bar{t}$ ($zb\bar{b}$) coupling. This suggests that in the heavy
case the $zt\bar{t}$ vertex
correction can be large even if $zb\bar{b}$ vertex correction is small
provided the mixing  and vetrex correction contributions
 are individually large and nearly equal.
  
b) Light case: In the light case $Z_1-Z_2$ mixing gives rise 
to the following [4] mass
eigenstates: $Z_1\approx Z_L-{sc^3\over x\cw}Z_H$
and $Z_2\approx Z_H+{sc^3\over x\cw}Z_L$. Hence the neutral gauge
boson mixing produces the following corrections to $\zl$ couplings

$$\delta g^{fm}_L= -{c^3s\over x}({c\over s}T_{3l}-{s\over c}T_{3h}).
\eqno(3)$$
In the light case the overall corrections to the LH $zb\bar{b}$
and $zt\bar{t}$ couplings are given by
$\delta g^b_L=
-{\xi^2 f^2_Q\over 2f^2}-{c^2s^2\over 2x}$ and 
$\delta g^t_L=-{\xi^2 f^2_Q\over 2f^2}+{c^2s^2\over 2x}$. We therefore find
that in the light case the effects due to mixing and  sideways ETC
induced vertex correction interfere constructively (destructively)
in $zb\bar{b}$ and $zt\bar{t}$ coupling.

\centerline{\bf II. $zt\bar{t}$ vertex correction in the heavy case}

The gauge boson mixing both in the charged and neutral sector modifies not
only the $zt\bar{t}$ and $zb\bar{b}$ couplings but also the SM prediction
to many other EW observables that are accessible at LEP 1. The two additional
parameters $s^2$ and ${1\over x}$ that determine the the gauge boson mixing
can be determined along with the usual standard model parameters from a
global fit to the precision EW data. In the heavy case for $s^2=.97$ and
$\alpha_s(M^2_z)\approx .115$ the best fit values of ${1\over x}$
and $\delta g^b_L$ obtanied by CST [4] are given by ${1\over x}=\approx
.0027\pm.0093$ and $\delta g^b_L\approx -.0064\pm .0074$.
We shall assume the above values for $s^2$ and
${1\over x}$, but treat $\delta g^b_L$
as a relatively free parameter since the LEP 1 value for $\rb$ has been
undergoing frequent changes. More precisely we shall  let $\delrb$
to assume the values .0022, .0044 .0066 and .0088 which correspond to 1\%,
2\%  3\% and 4\% deviations relative to the SM prediction for $\rb$.
It can be shown that in the heavy case the ETC correction to $\rb$
 is given by ${\delrb\over \rb}\approx .8973 \xi^2
  {\mt\over 4\pi f_Q}-.0047$ which can be 
  used to find the value of $ \xi^2{\mt\over 4\pi f_Q}$ for a given 
  $\delrb$. We find that for $\delrb =.0022, .0044, 
   .0066\  and\  .0088$\ \  ${\delftl\over \ftlsm}$ is given by
  -.0159, -.0240, -.0324 and -.0405 respectively. Here $\delftl$
  is the ETC induced correction to the LH form factor for $zt\bar{t}$
  vertex and $\ftlsm={1\over 2}-{2\over 3}s^2_w$ is its value in the SM. 
 Note that $\delftr=0$ both in heavy and light scenarios since: i)
 $Z_2$ does not couple to $t_R$ and therefore $Z_1-Z_2$ mixing does not
 renormalize the $Z_L t_R\bar {t}_R$ coupling and ii) the sideways 
 ETC induced four fermion  term $\bar{t}_R\gmmud T_R\bar {T}_R\gmmuu
 t_R$ after Fierz rearrangement does not contain any isospin triplet
technifermion current.
 
 \centerline{\bf III. $zt\bar{t}$ vertex correction in the light case.}

In the light case for $s^2=.97$ and $\alpha_s(M_z)=.115$, CST [4] found that
the best fit value for ${1\over x}$ lies in the unphysical region
(${1\over x}=-.17\pm.75$). However since the fit is fairly insensitive to
the value of ${1\over x}$, there is a substantial range of values
of ${1\over x}$ which provide a good fit to the data. Following
CST we shall choose ${1\over x}=.055$ in the light case.
Putting the mixing and vertex corrections together we find that 
${\delrb\over \rb}=.8973\xi^2{\mt\over 4\pi f_Q}+.7844(2.2879s^2+
2.3426c^2){c^2\over x}$, which can be used to find the value of
$\xi^2{\mt\over 4\pi f_Q}$ for a given $\delrb$.
We find that for $\delrb$= .0022, .0044, .0066 and .0088\ \  
${\delftl\over \ftlsm}$ is given by -.0035, -.0119, -.0200 and -.0281.

\centerline{\bf IV. Discussion of results}

We note the following features in the non-commuting ETC induced
correction to $zt\bar {t}$ couplings:

i) $\delftl$ is always very small in non-commuting ETC models. For 
$\delrb <.0088$ the correction to $F^t_L$ is at most
4.1\%(2.8\%) in the heavy (light) case. Since this is of the same order
as the usual SM corrections to $F^t_L$ it will be very hard to disentangle
one from the other at NLC. It should be noted that the calculation
of $\delftl$  does not depend on the value of $\mt$

ii) With increasing $\delrb$,\ \ \ $\delftl$ increases in magnitude
but the increase is very slow (almost inappreciable) in both light and
heavy cases.

iii) Non-commuting ETC interactions renormalize the $zt\bar {t}$ vertex
in such a way that the LH weak charge of the top quark always decreases in
magnitude.

v) For $\delrb = .0044$ we find that ${f_Q\over \xi^2}=497 Gev$
in the heavy case and 718 Gev in the light case. 
But $f_Q$ in the light case is expected to be less than its value in
 the heavy case. This can happen only if $\xi^2$ in the light case is 
  smaller than its value in the heavy case for non-commuting
 ETC models. 
 
The $zt\bar{t}$ couplings will be probed with good precision at NLC.
By including the $l^{\pm} +$ jets mode and making use of the angular
distribution of different polarization states of $t\bar {t}$, 
Ladinsky and Yuan [6] found that $F^t_L$ can be determined at NLC 500
to within
about 3\% (8\%) at 68\% (90\%) confidence limit. While $F^t_R$
can be known to within roughly 5\% (18\%).
The corrections to $F^t_L$ both in the heavy and light cases
being always less than or barely equal to the projected NLC sensitivity of
measuring it such corrections cannot be probed at NLC unless there is a
drastic improvement in its sensitivity. 
It is interesting to note that for $\delrb=.0044$, \ $\delftv=-\delfta
\approx-.0434$ for non-commuting ETC scenario. Whereas for $m^2_s=m^2_d$
and the same value for $\delrb$ \ $\delftv\approx -.179$ and 
$\delfta\approx -.068$ for diagonal ETC scenarios.
Non-commuting ETC scenarios are therefore expected to be much less
 constrained by $zt\bar {t}$ vertex correction than diagonal ETC scenarios.
 A 1 Tev machine can do better than
a 500 Gev machine in determining the corrections to $zt\bar {t}$ vertex
 because firstly the RR and LL events are 
suppressed relative to LR and RL events
 and secondly the top quark is boosted more which 
makes the determination of its momentum direction more accurate.

\centerline{\bf Conclusion}

In this article we have calculated the corrections to $zt\bar{t}$
couplings due to non-commuting ETC interactions. We find that the
for $\delrb < .0088$ \ \ $\delftl$ is always less than 4\% (2.8\%)
 in the heavy case (light case).
 Whereas $\delftr=0$ to the order to which we have worked.
  The non-commtuing ETC induced corrections
to $zt\bar{t}$ couplings  are therefore too small to be probed
with the projected NLC sensitivity. Further since these corrections are of
the same order as usual SM corrections it will be very hard to
disentangle one effect from the other. We therefore conclude
that it will be very hard to probe non-commuting ETC effects
by precision measurements of $zt\bar {t}$ couplings.

\centerline{\bf References}

\item{1.} R. S. Chivukula, K. Lane and A. G. Cohen, Nucl. Phys. B 343,
554 (1990); T. Appelquist, J. Terning and L. C. R. Wijewardhana, Phys.
Rev. D 44, 871 (1991).

\item{2.} R. S. Chivukula, S. B. Selipsky and E. H. Simmons, Phys. Rev.
Lett. 69, 575 (1992).

\item{3.} R. S. Chivukula, E. H. Simmons and J. Terning, Phys. Lett. B 331,
383 (1994).

\item{4.} R. S. Chivukula, E. H. Simmons and J. Terning, hep-ph/9506427,
BUHEP-95-19.

\item{5.} Talk presented by A. Blondel at the Warsaw ICHEP, July 1996.

\item{6.} G. A. Ladinsky and C. P. Yuan, Phys. Rev. D 49, 4415 (1994).

\end